\documentclass[conference]{IEEEtran}
\IEEEoverridecommandlockouts
\usepackage{cite}
\usepackage{amsmath,amssymb,amsfonts}
\usepackage{algorithmic}
\usepackage{graphicx}
\usepackage{textcomp}
\usepackage{xcolor}
\usepackage{listings}
\usepackage{booktabs}
\usepackage{float}
\usepackage{stfloats}
\usepackage{caption} 

\def\BibTeX{{\rm B\kern-.05em{\sc i\kern-.025em b}\kern-.08em
    T\kern-.1667em\lower.7ex\hbox{E}\kern-.125emX}}
    
\newcommand{\SPR}{4\textsuperscript{th} Gen Intel Xeon}
\newcommand{\SPRH}{Intel Xeon Processors codenamed Sapphire Rapids with HBM}
\usepackage{array}
\newcommand{\PreserveBackslash}[1]{\let\temp=\\#1\let\\=\temp}
\newcolumntype{C}[1]{>{\PreserveBackslash\centering}p{#1}}

\begin{document}

\title{Early Performance Results on 4th Gen Intel(R)  Xeon (R) Scalable Processors with DDR and Intel(R) Xeon(R) processors, codenamed Sapphire Rapids with HBM 
\thanks{This work was supported by the U.S. Department of Energy through the Los Alamos National Laboratory (LANL). 
LANL is operated by Triad National Security, LLC, for the National Nuclear Security Administration of U.S. 
Department of Energy (Contract No. 89233218CNA000001). LA-UR-22-31875 Approved for public release; distribution is unlimited.}
}

\author{

\IEEEauthorblockN{$^\ddag$Galen M. Shipman, Sriram Swaminarayan\\}
\IEEEauthorblockA{\textit{Computer, Computational, and Statistical Sciences} \\
\textit{Los Alamos National Laboratory}\\
Los Alamos NM, USA \\
gshipman@lanl.gov, sriam@lanl.gov} \\
\and
\IEEEauthorblockN{Gary Grider, Jim Lujan\\}
\IEEEauthorblockA{\textit{High Performance Computing} \\
\textit{Los Alamos National Laboratory}\\
Los Alamos NM, USA \\
ggrider@lanl.gov, jewel@lanl.gov}\\
\and
\IEEEauthorblockN{R. Joseph Zerr\\}
\IEEEauthorblockA{\textit{X Computational Physics} \\
\textit{Los Alamos National Laboratory}\\
Los Alamos NM, USA \\
rzerr@lanl.gov}

}

\maketitle

\begin{abstract}

The Crossroads supercomputer was designed to simulate some of the most complex physical devices in the world. These simulations routinely require 1/2 petabyte or more of system memory running on thousands of compute nodes for months at a time on the most powerful supercomputers. Improvements in time to solutions for these workloads can have major impact on our mission capabilities. In this paper we present early results of representative application workloads on \SPR~and \SPRH. These results demonstrate an extremely promising 8.57x improvement (node to node) over our prior generation Intel Broadwell (BDW) based HPC systems. No code modifications were required to achieve this speedup, providing a compelling path forward toward major reductions in time to solution and the complexity of physical systems that can be simulated in the future. 
 
\end{abstract}


\section{Introduction}

Los Alamos National Laboratory (LANL) has a long history in advancing the state-of-the-art in High Performance Computing for complex multi-physics simulations. The Crossroads supercomputer scheduled for deployment in 2023 will be the latest in a series of HPC systems developed in collaboration with Intel and HPE/Cray. Crossroads will include  4\textsuperscript{th} Gen Intel Xeon Scalable processors, formerly codenamed Sapphire Rapids, based on the Intel 7 process node and the novel Embedded Multi-Die Interconnect Bridge (EMIB) chiplet integration technology. This CPU promises to deliver significant improvements in simulation performance with little if any code modifications. As many of our codes are significantly memory bound, LANL worked with Intel to develop a Intel Xeon Scalable Processor with High Bandwidth Memory (HBM) 2e integrated with EMIB technology.  

Initial experience on pre-production \SPR~with Double Data Rate (DDR) memory shows up to a 5.83x performance improvement in our multi-physics simulations compared to our current Broadwell based production systems. Pre-production \SPRH~shows up to a 8.57x performance improvement. These improvements were achieved with no code changes, simply a recompile of the code to target the new microarchitecture.

\section{\SPR~and \SPRH~Architectures}\label{sec:spr_architecture}

\SPR~is the latest generation Xeon server processor based on Golden Cove microarchitecture (using Intel 7 process technology). A number of significant innovations from Intel are present in the processor including EMIB, acceleration engines, CXL, and bandwidth improvements in Ultra Path Interconnect. Microarchitecture and instructions per cycle (IPC) improvements include a 2x increase in decode length, 50\% increase in decode width, a 2.4x increase in branch target size, ~1.8x increase in the $\mu$op cache, a larger number of execution ports and and a ~1.5x increased reorder buffer. These microarchitecture changes improve the IPC rate and enable increased parallelism within the CPU. 

\SPR~is also the first CPU to incorporate Intel's EMIB technology providing a more scalable path to further performance improvements and efficiency relative to traditional monolithic designs that are limited to a single reticle. This ability to provide tight integration beyond the single reticle limit is an important advancement, and it has other benefits including the potential for higher yield and the ability to integrate disparate processor and memory technologies in a single package. \SPR~is composed of four ~400$mm^2$ tiles each with golden cove cores with 48KB instruction and 32KB data caches. L2 is private and is 2048KB. Our system configurations are either dual socket \SPR~or dual socket \SPRH~each with 66 physical cores for a total of 112 physical cores per node. \SPR~configuration has 8 channels of 16 GB DDR-5 operating at 4800 MT/s per socket for a total of 256GB of memory per 2 CPU node. \SPRH~configuration has 4 banks of 8 high 16 Gbit HBM2e  operating at 3200 MT/s per socket for a total of 128 GB of memory per 2 CPU node.

\section{Multi-Physics Simulation Codes} \label{sec:applications} 

LANL develops and maintains a suite of multi-physics simulation codes used to support our national security and science mission. These codes are used daily by hundreds of scientists and engineers to answer questions of national importance. Significant effort in validation and verification of these codes underwrite their capabilities and provide confidence in their use to simulate complex physical devices. 
As a result of this work, scientists are able to to explore complex issues in  Inertial Confinement Fusion~\cite{grinstein2022transition}, a regime of physics that is only achievable at a single experimental device in the world, the National Ignition Facility (NIF). The ability to simulate multi-material reactive burn in high pressure regimes has enabled the simulation~\cite{coleman2020modeling} of the overdriven detonation states in the triple point overdrive experiment conducted by LANL.

\subsection{xRAGE} \label{sec:xrage} 

xRAGE~\cite{gittings2008rage}  is an Eulerian radiation/hydrodynamics code under active development at LANL. xRAGE (Radiation Adaptive Grid Eulerian) supports simulations in 1D, 2D, and 3D,  multiple materials, and coupling of radiation diffusion and hydrodynamics. Adaptive Mesh Refinement (AMR) of unit aspect ratio cell volumes supports large dynamic ranges of state space (centimeters to kilometers, microns to meters) with refinement and derefinement of adjacent $2^d$ cells (2 in 1D, 4 in 2D, and 8 in 3D) each timestep.  

Most simulations using xRAGE are DRAM bandwidth bound with relatively low arithmetic intensity. This may surprise some readers as conventional wisdom is that HPC applications are largely FLOP bound, but prior analysis has shown most subroutines in xRAGE simulations are completely or significantly memory bound with an arithmetic intensity of 0.001 to 0.2 FLOP/byte using Intel roofline~\cite{marques2017performance} analysis. Analysis using  BYFL~\cite{pakin2013hardware} "software based performance counters" illustrated in Table~\ref{tab:byfl} details the types of 
instructions and the percentage issued from a representative xRAGE simulation. Sparse memory operations result in load/store operations coupled with integer and array indexing operations and represent 56\% of all instructions issued in this simulation as detailed in Table~\ref{tab:byfl} in blue font.  Fine- and coarse-grained branching can be seen in the relatively high (25) percent of conditional/branching instructions issued in this simulation, as detailed in Table~\ref{tab:byfl} in red font. Floating point operation represent only 6\% of instructions issued.

\begin{table}
  \centering
  \begin{tabular}{ | l | r | r | }
    \hline
    \textbf{Instruction} & \textbf{Count} & \textbf{Percentage} \\
    \hline\hline
    \color{blue} Load & \color{blue} 6,775,030,849 & \color{blue} 18\% \\ 
    \hline
    \color{red} Branching &  \color{red} 6,063,697,707 &  \color{red} 16\% \\  
    \hline
    \color{blue} Integer Add & \color{blue}  5,334,155,682 & \color{blue}  15\% \\
    \hline
    \color{blue} Array Indexing & \color{blue}  4,855,537,532 & \color{blue}  13\% \\
    \hline
    \color{red} Conditional & \color{red} 3,299,248,274 & \color{red} 9\% \\
    \hline
    \color{blue} Store & \color{blue}  2,599,966,427 & \color{blue}  7\% \\
    \hline
    Type cast & 1,959,938,043 & 5\%\\
    \hline
    Sign extension & 1,541,094,404 & 4\% \\
    \hline
    Stack frame allocation & 1,221,694,311 &  3\%\\
    \hline
    FP multiplication & 1,171,615,897 & 3\% \\
    \hline
    FP comparison & 1,141,415,386 & 3\% \\
    \hline
    \color{blue} INT multiplication & \color{blue} 991,524,374 & \color{blue} 3\%  \\
    \hline
  \end{tabular}
  \caption{Instruction breakout by type in a representative simulation~\cite{pakin2013hardware}. Red denotes operations attributable   to branching (25\%). Blue denotes operations attributable to sparse data structures (56\%). }
  \label{tab:byfl}
\end{table}

This and deeper analysis~\cite{shipman2022memory} motivated a focused effort on addressing the memory wall in the Crossroads architecture.

\subsection{FLAG} \label{sec:flag} 
FLAG is a  Lagrange-based radiation/hydrodynamics code that supports multiple mesh optimization strategies in 1, 2, and 3D with support for fully unstructured (polyhedral) grids. FLAG supports arbitrary Lagrangian-Eulerian (ALE) and adaptive mesh refinement methods to handle mesh distortions / tearing and multiple levels of resolution. 

Similar to xRAGE simulations, most simulations using FLAG are also DRAM bandwidth bound with relatively low arithmetic intensity. Memory access patterns in FLAG are often sparse, which the Pennant~\cite{ferenbaugh2015pennant} proxy application attempts to capture. Analysis of Pennant, done as part of the Spatter~\cite{lavin2020evaluating} work, shows a significant number of scatter/gather patterns with non-unit strides. Some evidence gathered with gem5 simulations~\cite{shipman2022memory} further suggests that loaded latency~\cite{lavin2020evaluating} can also impact the performance of these simulations significantly.

The performance characteristics of these two codes, xRAGE and FLAG, are representative of a large number of LANL codes that make use of similar data structures. Improving the performance of these codes through architecture innovation has a broad impact to wide range of codes and their applications at LANL and therefore motivates our focus here. While both codes are capable of simulating hydrodynamics coupled with radiation transport this early study focuses exclusively on multi-material hydrodynamics test problems.

\section{Experiments and Results}\label{sec:experiments}

All experiments were conducted on Linux based dual-socket nodes. \SPR~and \SPRH~nodes were configured as detailed above in section~\ref{sec:spr_architecture}. Intel Broadwell nodes were configured with E5-2695 V4 with 4 channels per socket of DDR4 operating at 2400 MT/s. Linux processes were bound to individual cores and memory was affinitized to the core to eliminate QPI traffic. All benchmarks were conducted without hyperthreading and without over-subscription of cores. For node-to-node comparisons, benchmark problems were limited to the 128 GB available on our Intel Broadwell and \SPRH. 

\begin{figure*}[t]
 \begin{center}
\includegraphics[width=16cm]{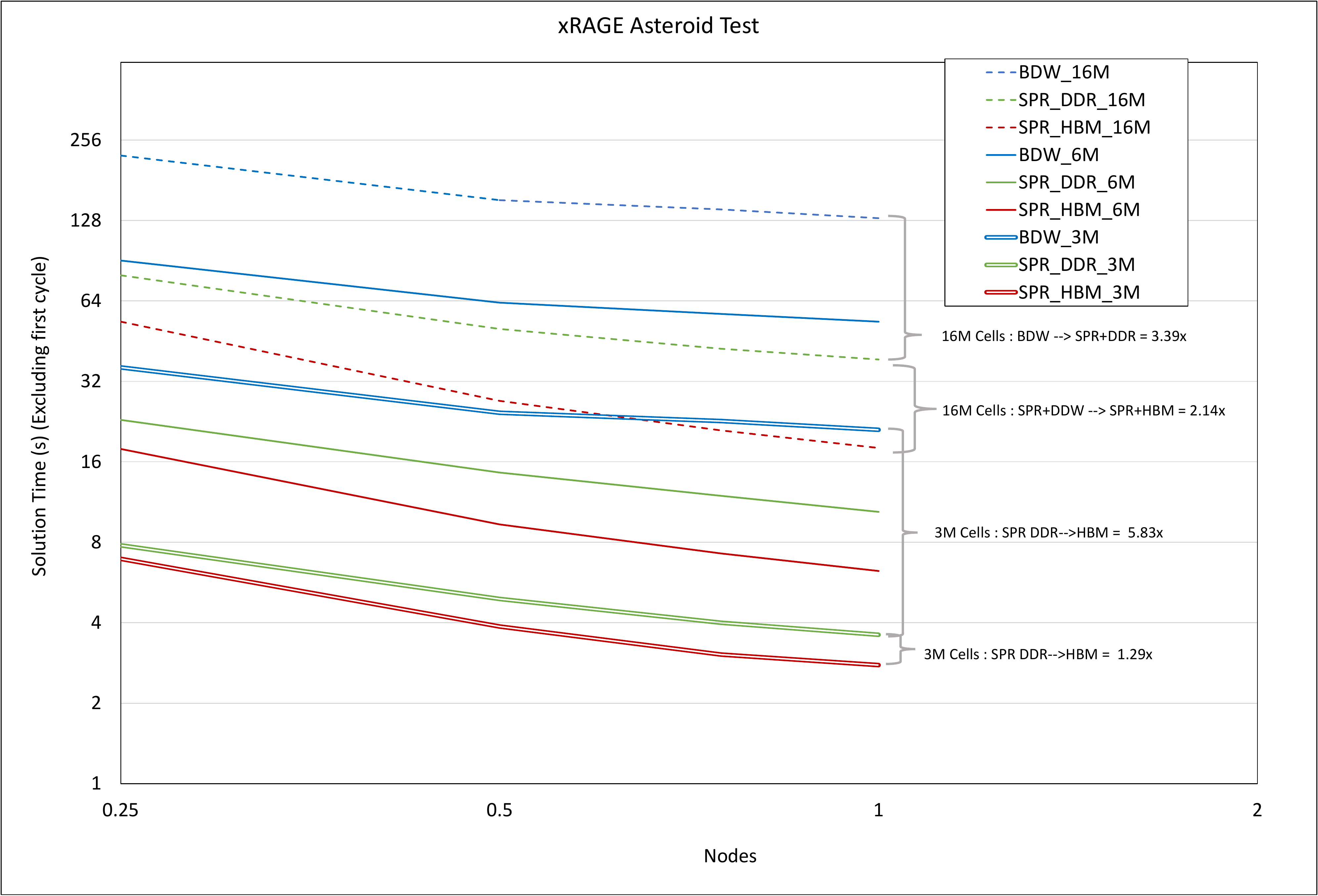}
\caption{Strong scaling of xRAGE Asteroid test problem at 1/4, 1/2, and a full dual socket Broadwell (BDW), \SPR~(SPR+DDR), and \SPRH~(SPR+HBM) nodes}
\label{fig:eap-strong}
\end{center}
\end{figure*}

The xRAGE code was used to benchmark an Asteroid impact problem using a configuration similar to those described in the literature~\cite{gisler2004two}. Three problem sizes are given, one using 3 million mesh cells and a resident set size of up to 46GB of system memory (Asteroid\_3M), another using 6 million mesh cells and a resident set size of up to 53GB of system memory (Asteroid\_6M), and a third using 16 million mesh cells and a resident set size of up to 104GB (Asteroid\_16M).
Table~\ref{tab:xragedata} details the time to solution improvements of Asteroid simulations using xRAGE on \SPR~relative to BDW DDR and \SPRH~relative to \SPR. Overall performance of both \SPR~ and \SPRH~is significantly higher than that achieved by BDW DDR. The smaller benchmark problem Asteroid\_3M is 5.83x faster on \SPR~ relative to BDW DDR and 7.58x faster on \SPRH~relative to BDW DDR. Larger problems such as Asteroid\_6M and Asteroid\_16M result in speedups of 5.14x and 3.39x on \SPR~ and 8.57x and 7.25x on \SPRH~ demonstrating the significant performance improvements resulting in the HBM memory bandwidth. Strong scaling plots from 1/4, 1/2 and a full dual socket Broadwell, \SPR~and \SPRH~are illustrated in Figure~\ref{fig:eap-strong}. 

\begin{table}[h]
    \centering
    \begin{tabular}{ | r | l | l | l | }
        \hline &  \multicolumn{3}{c}{Relative time to solution} \\
        \hline
        Benchmark  &  BDW+DDR / &  BDW+DDR / & SPR+DDR / \\
             & SPR+DDR  &  SPR+HBM / & SPR+HBM \\
        \hline
        Asteroid\_3M & 5.83 & 7.58 & 1.29\\
        \hline
        Asteroid\_6M & 5.14 & 8.57 & 1.66 \\
        \hline
          Asteroid\_16M & 3.39 & 7.25 & 2.14  \\
        \hline
    \end{tabular}
    \caption{Results of three simulations run using xRAGE utilizing 36 cores on Intel Broadwell (BDW) and 112 cores on \SPR~(SPR+DDR) and \SPRH~(SPR+HBM) nodes.}
    \label{tab:xragedata}
\end{table}

\begin{figure*}[t]
\begin{center}
\includegraphics[width=15cm]{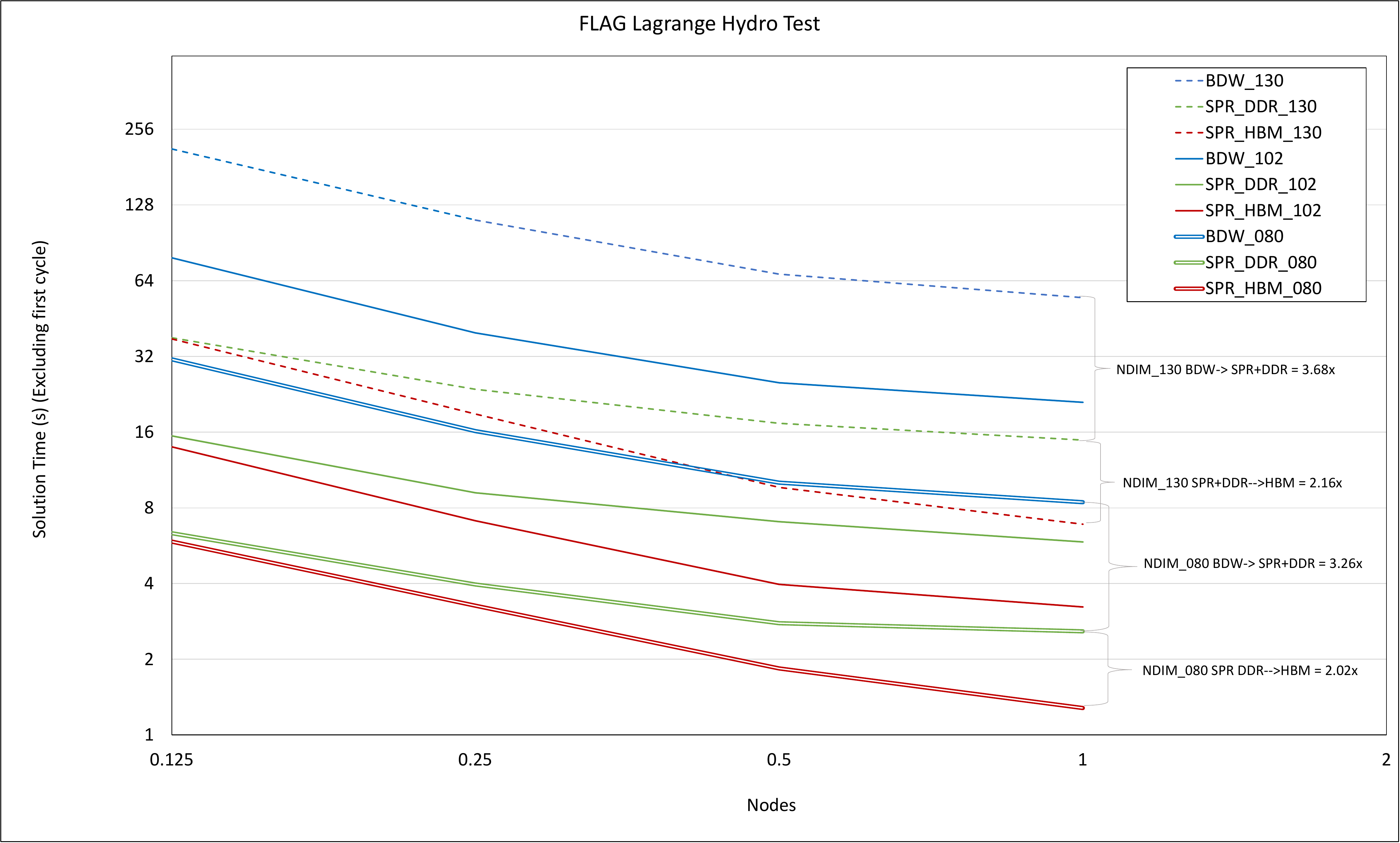}
\caption{Strong scaling of the FLAG Lagrange Hydro SOD test problem at 1/4, 1/2, and a full dual socket Broadwell (BDW), \SPR~(SPR+DDR), and \SPRH~(SPR+HBM) nodes}
\label{fig:lap-lag-strong}
\end{center}
\end{figure*}
\begin{figure*}[t]
\begin{center}
\includegraphics[width=15cm]{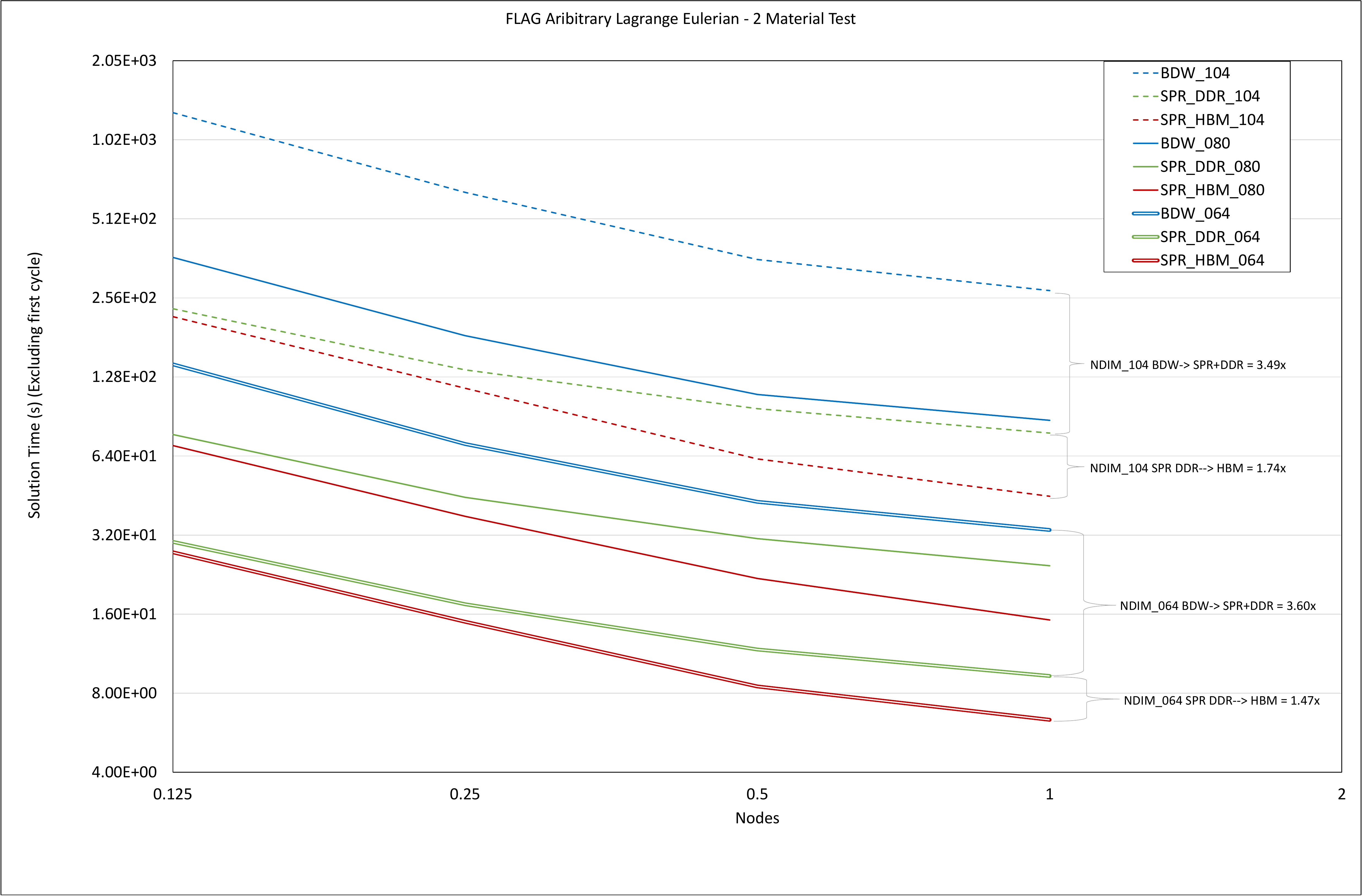}
\caption{Strong scaling  of the FLAG  ALE Hydro SOD test problem at 1/4, 1/2, and a full dual socket Broadwell (BDW), \SPR~(SPR+DDR), and \SPRH~(SPR+HBM) nodes}
\label{fig:lap-ale-strong}
\end{center}
\end{figure*}

 The FLAG code was used to benchmark a Sod shock tube test problem described in~\cite{burton2013compendium}. Two mesh management strategies are used, one using Lagrange hydrodynamics where the mesh "moves" during the simulation, labelled "LH". The second uses ALE in which a Lagrange step is calculated and the material is advected (Eulerian step) to the fixed mesh labelled "ALE" for ALE with 2 materials. Table~\ref{tab:flagdata} details the time to solution improvements of Sod shock tube simulations using FLAG on  \SPR~ relative to BDW DDR and \SPRH~relative to \SPR. Performance of Lagrange Hydro with a problem size of $130^3$ (3D mesh) is 3.68x higher on \SPR~ relative to BDW DDR and is 7.94x higher on \SPRH. Smaller problem sizes of $102^3$ and $80^3$ using LH show somewhat smaller but still quite respectable performance improvements on \SPR~ and \SPRH. The smaller problem sizes likely have a negligible cache effect here as they are still relatively large. Performance of ALE Hydro with a problem size of $104^3$ (3D mesh) is 3.49x higher on \SPR~relative to BDW DDR and is 6.07x higher on \SPRH. Smaller problem sizes of $80^3$ and $64^3$ using ALE show a bit larger performance improvements on \SPR~(3.58x and 3.6x) and a bit lower on \SPRH~(5.76x and 5.28x).  ALE requires significantly more instructions relative to LH for each hydrodynamics cycle which likely accounts for some of these differences. Further analysis using roofline modeling and frontend / backend analysis to determine potential contributors to this result is warranted. Strong scaling plots from 1/4, 1/2 and a full dual socket Broadwell, \SPR~and \SPRH~for the FLAG Lagrange Hydro SOD test problem are illustrated in Figure~\ref{fig:lap-lag-strong}. Strong scaling results for the FLAG ALE SOD Hydro test problem are illustrated in Figure~\ref{fig:lap-ale-strong}.

\begin{table}[h]
    \centering
    \begin{tabular}{ | r | l | l | l | }
        \hline &  \multicolumn{3}{c}{Relative time to solution} \\
        \hline
        Benchmark  &  BDW+DDR / &  BDW+DDR / & SPR+DDR / \\
             & SPR+DDR  &  SPR+HBM & SPR+HBM \\
        \hline
        LH\_NDIM\_80 & 3.26 & 6.57 & 2.02 \\
        \hline
        LH\_NDIM\_102 & 3.59 & 6.51 & 1.81 \\
        \hline
        LH\_NDIM\_130 & 3.68 & 7.94 & 2.16 \\
        \hline
        ALE\_NDIM\_64 & 3.60 & 5.28 & 1.47\\
        \hline
        ALE\_NDIM\_80 & 3.58 & 5.76 & 1.61  \\
        \hline
         ALE\_NDIM\_104 & 3.49 & 6.07 & 1.74 \\
        \hline
    \end{tabular}
    \caption{Results of six simulations run using FLAG on two different test problems utilizing 36 cores on Intel Broadwell (BDW) and 112 cores on \SPR~(SPR+DDR) and \SPRH~(SPR+HBM) nodes.}
    \label{tab:flagdata}
\end{table}

\section{Conclusions}\label{sec:conclusions} 

Multi-physics codes designed to simulate large state and phase spaces with many materials are often memory bandwidth  rather than FLOP bound. Addressing this bottleneck has proven quite challenging as many architectures have continued to grow in raw FLOP performance while memory bandwidth increases have been more modest. To address this, LANL has worked closely with Intel to integrate high-bandwidth memory in the \SPRH~as part of the Crossroads Supercomputer design. Early experience with pre-production silicon shows extremely promising results with performance improvements up to 8.57x over current HPC systems at LANL. While other technologies have at times required significant refactoring or complete rewrites of major portions of a code, a simple recompile was all that was necessary when porting to \SPRH. These results demonstrate a fruitful path towards efficient computing technologies in the future for LANL's most challenging applications.  

\section*{Acknowledgment}

The authors thank members of the HPC division, the Darwin testbed team, the Performance Engineering Team,  the Applications Readiness Team, the Eulerian Applications Project, and the Lagrangian Applications Project at Los Alamos National Laboratory. 

\bibliographystyle{unsrt}
\bibliography{intel-spr}

\begin{thebibliography}{10}

\bibitem{grinstein2022transition}
Fernando Grinstein, Vincent Chiravalle, and Brian Haines.
\newblock Transition and multiphysics in 3d icf capsule implosions.
\newblock {\em Bulletin of the American Physical Society}, 2022.

\bibitem{coleman2020modeling}
Adam~C Coleman, Carl~E Johnson, and Matthew~M Biss.
\newblock Modeling the lanl triple point overdrive experiment in the flag
  hydrocode.
\newblock In {\em AIP Conference Proceedings}, volume 2272, page 030007. AIP
  Publishing LLC, 2020.

\bibitem{gittings2008rage}
Michael Gittings, Robert Weaver, Michael Clover, Thomas Betlach, Nelson Byrne,
  Robert Coker, Edward Dendy, Robert Hueckstaedt, Kim New, W~Rob Oakes, et~al.
\newblock The rage radiation-hydrodynamic code.
\newblock {\em Computational Science \& Discovery}, 1(1):015005, 2008.

\bibitem{marques2017performance}
Diogo Marques, Helder Duarte, Aleksandar Ilic, Leonel Sousa, Roman Belenov,
  Philippe Thierry, and Zakhar~A Matveev.
\newblock Performance analysis with cache-aware roofline model in intel
  advisor.
\newblock In {\em 2017 International Conference on High Performance Computing
  \& Simulation (HPCS)}, pages 898--907. IEEE, 2017.

\bibitem{pakin2013hardware}
Scott Pakin and Patrick McCormick.
\newblock Hardware-independent application characterization.
\newblock In {\em 2013 IEEE International Symposium on Workload
  Characterization (IISWC)}, pages 111--112. IEEE, 2013.

\bibitem{shipman2022memory}
Galen Shipman, Jered Dominguez-Trujillo, Kevin Sheridan, and Sriram
  Swaminarayan.
\newblock Assessing the memory wall in complex codes.
\newblock In {\em 2022 IEEE/ACM Workshop on Memory Centric High Performance
  Computing (MCHPC)}, 2022.

\bibitem{ferenbaugh2015pennant}
Charles~R Ferenbaugh.
\newblock Pennant: an unstructured mesh mini-app for advanced architecture
  research.
\newblock {\em Concurrency and Computation: Practice and Experience},
  27(17):4555--4572, 2015.

\bibitem{lavin2020evaluating}
Patrick Lavin, Jeffrey Young, Richard Vuduc, Jason Riedy, Aaron Vose, and
  Daniel Ernst.
\newblock Evaluating gather and scatter performance on cpus and gpus.
\newblock In {\em The International Symposium on Memory Systems}, pages
  209--222, 2020.

\bibitem{gisler2004two}
Galen~R Gisler, Robert~P Weaver, Charles~L Mader, and Michael~L Gittings.
\newblock Two-and three-dimensional asteroid impact simulations.
\newblock {\em Computing in Science \& Engineering}, 6(3):46--55, 2004.

\bibitem{burton2013compendium}
DE~Burton, TC~Carney, NR~Morgan, and MA~Kenamond.
\newblock Compendium of cch and xale test problems.
\newblock {\em Los Alamos National Laboratory}, 2013.

\end{thebibliography}

\end{document}